\documentclass[amsmath, amsfonts, superscriptaddress, showpacs, twocolumn,prb]{revtex4}
\usepackage{graphicx}
\usepackage{epsfig}
\usepackage{bm}
\usepackage{dcolumn}
\usepackage{amsmath}
\usepackage{amssymb}

\renewcommand{\vec}[1]{{\bm #1}}
\renewcommand{\Im}{\mathrm{Im}}
\renewcommand{\Re}{\mathrm{Re}}

\begin{document}

\title{Raman scattering as a probe of nematic correlations}

\author{M.~Khodas}
\affiliation{Department of Physics and Astronomy, University of Iowa, Iowa City, Iowa 52242, USA}
\affiliation{Racah Institute of Physics, Hebrew University of Jerusalem, Jerusalem 91904, Israel}

\author{A.~Levchenko}
\affiliation{Department of Physics, University of Wisconsin-Madison, Madison, Wisconsin 53706, USA}
\affiliation{Department of Physics and Astronomy, Michigan State University, East Lansing, Michigan 48824, USA}

\begin{abstract}
We use the symmetry-constrained low-energy effective Hamiltonian of iron-based superconductors to study the Raman scattering in the normal state of underdoped iron-based superconductors.
The incoming and scattered Raman photons couple directly to orbital fluctuations and indirectly to the spin fluctuations. We computed both couplings within the same low-energy model.
The symmetry-constrained Hamiltonian yields the coupling between the orbital and spin fluctuations of only the same symmetry type.  Attraction in the $B_{2g}$ symmetry channel was assumed for the system to develop the subleading instability towards the discrete in-plane rotational symmetry breaking, referred to as Ising nematic transition. We find that upon approaching this instability, the Raman spectral function develops a quasielastic peak as a function of energy transferred by photons to the crystal. We attribute this low-energy $B_{2g}$ scattering to the critical slowdown associated with the build up of nematic correlations.
\end{abstract}

\date{June 4, 2015}


\maketitle

\section{Introduction}

The discovery of iron-based superconductors (FeSCs) opened new avenues in the research of strongly correlated systems.~\cite{Mazin2010,Paglione2010,Johnston2010,Stewart2011,Basov2011} Despite the diversity in crystallographic structure and chemical composition, all the FeSCs share several generic trends. FeSCs are multiband and multipocket materials. According to angle-resolved photoemission spectroscopy (ARPES) the Fermi surface (FS) contains two or three hole pockets at the center of the Brillouin zone, the $\Gamma$ point, and two electron pockets centered at $(\pi,\pi)$, the $M$ point in two-iron-atom unit-cell notations. The underdoped compounds undergo structural tetragonal to orthorhombic transition at the temperature $T_s$ followed by or coincident with the spin density wave (SDW) transition at $T_{SDW}$. The superconductivity sets in when the magnetism is suppressed by doping~\cite{Kamihara2006,Kamihara2008,Chen2008,Sefat2008,Chu2009} or pressure~\cite{Takahashi2008,Hamlin2008}. 

The interplay between the magnetism and superconductivity is manifest in the weak-coupling renormalization-group analysis of competing instabilities.~\cite{Chubukov2008,Chubukov2009b,Chubukov2012} The interaction amplitude in the spin-density-wave channel is renormalized in a way similar to that for the usual renormalization in the particle-particle channel that normally leads to Cooper instability.  Above the Fermi energy $E_F$ the two channels affect each other. As a result, the interpocket pairing interaction is enhanced by the spin fluctuations, which were suggested to drive the unconventional $s^{\pm}$ superconductivity with the order parameter changing sign between the electron and hole FSs. In this picture low (high) doping makes the magnetic (Cooper) instability a winner in a competition at energies below $E_F$. It follows that the proximity of the magnetic and superconducting phases on a phase diagram is not accidental. Hence, the understanding of magnetic and structural transitions is instrumental for the description of the superconductivity.

Most commonly, at the magnetic transition the continuous $O(3)$ symmetry and the discrete time-reversal symmetry are broken. In the FeSCs the spin alignment is magnetic along one direction and antiferromagnetic in the orthogonal direction. Such stripe magnetization lowers, therefore, the discrete $C_4$ rotational symmetry of the lattice down to $C_2$. The possibility of breaking the $C_4$ symmetry without breaking the spin  $O(3)$ and time-reversal symmetry was studied in the context of the structural transition, and the corresponding transition was referred to as being Ising nematic.~\cite{Fernandes2012}
In this picture, below $T_s$ the spin-correlation length increases in one of the symmetry directions and decreases in the other, and the magnetisms sets in with little or no delay.

The prevailing scenario of the structural transition is electronic. Specific to FeSCs is a rather high degree of $ab$ anisotropy in the electronic properties. The resistivity anisotropy $\rho_b/\rho_a$ in cobalt-doped BaFe$_2$As$_2$ is reported \cite{Chu2010} to reach values as high as 2 for cobalt concentration $x \approx 0.03$, whereas the maximal orthorhombic distortion for the parent material, $x=0$, is only about 0.36$\%$.  Moreover, in {\it strain}-controlled samples the derivative of $(\rho_b-\rho_a)/(\rho_b+\rho_a)$ with respect to the strain shows a divergence at the interpolated mean-field temperature $T^*$ = 116K for the parent compound.~ \cite{Chu2012} The $T^*$ so obtained is only 22K lower than the actual transition temperature, $T_s $= 138K. The relatively small difference $T_s - T^*$ is due to the lattice fluctuations being suppressed under the conditions of fixed strain. Likewise, the optical reflectivity is nearly divergent at the nematic transition.~\cite{Mirri2014} All these findings are indicative of the dominance of the electronic degrees of freedom in the nematic transition.

It is, in general, hard to disentangle different electronic fluctuation channels breaking the same symmetry. 
At present the dominance of either charge or spin degrees of freedom in driving the structural  transition is not settled. There are two schools of thought as to the origin of electronic nematicity.~ \cite{Fernandes2014} In the orbital nematicity scenario the difference in populations  $n_{Xz} - n_{Yz}$  of the $d_{Xz}$ and $d_{Yz}$ iron orbitals is believed to be the primary cause of the nematic transition.~ \cite{Kruger2009,Lv2009,Lee2009a,Kontani2012,Onari2012} In another scenario it is the spin that drives the nematic transition.~\cite{Xu2008,Fernandes2014} Let  $\vec{m}_{1,2}$ be the two staggered (antiferromagnetic) magnetizations on the even and odd iron sublattices, respectively. The nematic transition occurs when the two spin sublattices lock, $\langle \vec{m}_{1} \cdot \vec{m}_2 \rangle \neq 0$.~\cite{Fernandes2010b} The two alternatives are the positive and negative $\langle \vec{m}_{1} \cdot \vec{m}_2 \rangle$, resulting in two orthogonal stripe magnetizations, $\vec{\Delta}^{X,Y}=\vec{m}_1 \pm \vec{m}_2$. These are the spin arrangements ferromagnetic in the $X$($Y$) direction and antiferromagnetic in the $Y$($X$) direction in an Fe-only lattice. The magnetic perspective is supported by the NMR data showing a low-$T$ Curie-Weiss-like upturn of a spin-lattice relaxation rate $1/T_1T$,~\cite{Ning2010,Nakai2010} and by the scaling between the magnetic fluctuations and softening of the elastic shear modulus at the structural transition.~\cite{Fernandes2013b}

In this paper we do not attempt to resolve the above controversy, but rather explore the consequences of the nematic fluctuations as observed in recent Raman experiments.~\cite{Gallais2013,Zhang2014,Thorsmolle2014} Even though the region of the phase diagram contained between $T_s$ and $T_{SDW}$ is either absent or quite tiny, the dynamical nematic fluctuations revealed by Raman spectroscopy kick in far into the paramagnetic phase up to room temperatures. The Raman spectroscopy is essentially a dynamic probe of electronic correlations of prescribed symmetry.~\cite{Klein1984,Devereaux2007} The photon scattered inelastically leaves some of its energy with the crystal. Selection rules fix the symmetry of the excitation, while the energy difference between the incoming and scattered photons, the so-called Raman shift, determines the energy of the electronic excitations.

\section{Raman response in the four-band model}

\subsection{Band structure model}

\begin{figure}[t!]
  \includegraphics[width=9cm]{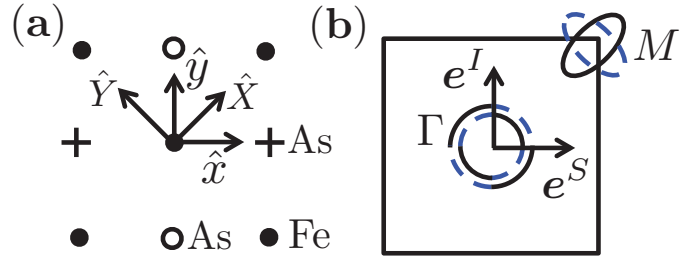}\\
  \caption{(color online) (a) The unit cell of a quasi-two-dimensional FeSC contains two iron atoms and two pnictogen atoms such as As (or a chalcogen atom such as Se). The atoms above and below the iron layer are denoted by crosses and by circles respectively. The basis vectors of the iron-only lattice are denoted by $\hat{X}$ and $\hat{Y}$. The vectors $\hat{x}$ and $\hat{y}$ are chosen as a basis vectors of the two-iron unit cell lattice. (b) The two-iron Brillouin zone. The $\Gamma$ point hosts two hole pockets and the $M$ points hosts two electron pockets. The solid (black) and dashed (blue) lines denote the $d_{Xz}$ and $d_{Yz}$ orbital contents respectively. The admixture of the $d_{XY}$ orbital at the outer parts of the crossed Fermi pockets at $M$ is neglected. The polarization vectors $\vec{e}^I$ and $\vec{e}^S$ for the $B_{2g}$ Raman configuration are shown.
}\label{fig:polarization}
\end{figure}

In this section we discuss the phenomenological four-band model based on the work of Cvetkovic and Vafek.~\cite{Cvetkovic2013} In this model constructed using the method of invariants due to Luttinger~\cite{Luttinger1956} the interaction of electrons with light is easily obtained using the standard gauge-invariant minimal-coupling procedure~\cite{Devereaux2007}. Here we neglect the coupling between the different layers and consider the crystal structure to be quasi-two-dimensional, [see Fig.~\ref{fig:polarization}(a)]. Generically, in FeSCs each layer contains the iron atoms forming a simple square lattice with the basis unit vectors $\hat{X}$ and $\hat{Y}$. The pnictogen or chalcogen atoms form the checkerboard with even and odd sublattices above and below the iron layer. Above the SDW transition the unit cell contains two iron atoms with the basis denoted by $\hat{x}$ and $\hat{y}$.

It is sufficient to consider a slightly simplified version of the model \cite{Cvetkovic2013} in which the four-dimensional effective Hamiltonian describing the $M$ point is replaced with the two-dimensional one and the remaining electronic bands that are not crossing the Fermi level are discarded. We write, for the quadratic part of the Hamiltonian,
\begin{align}\label{Hfree}
\mathcal{H} =  \sum_{\vec{k},\alpha}\sum_{i,j=1,2}
c_{i,\vec{k}\alpha}^{\dag}  \mathcal{H}^{\Gamma}_{\vec{k};i,j} c_{j,\vec{k}\alpha}
+
f_{i,\vec{k}\alpha}^{\dag}  \mathcal{H}^{M}_{\vec{k};i,j} f_{j,\vec{k}\alpha}\, ,
\end{align}
where $c_{i,\vec{k}\alpha}^{\dag}$ ($f_{i,\vec{k}\alpha}^{\dag}$) creates a hole (an electron) in a state with spin index $\alpha$ and momentum $\vec{k}$ counted from the $\Gamma$ ($M$) point. The index $i=1$ ($i=2$) refers to the $d_{Xz}$ ($d_{Yz}$) orbital content. For that reason the Hamiltonian \eqref{Hfree} refers to the orbital basis and reads
\begin{align}\label{HGamma}
\mathcal{H}^{\Gamma}_{\vec{k}} = 
\begin{bmatrix}
\epsilon_{\Gamma} + \frac{k^2}{2 m_{\Gamma}} +  a  k_x k_y & \frac{c}{2}(k_x^2 - k_y^2) \\
\frac{c}{2}(k_x^2 - k_y^2) & \epsilon_{\Gamma} + \frac{k^2}{2 m_{\Gamma}} -   a k_x k_y 
\end{bmatrix}
\end{align}
for holes and 
\begin{align}\label{HM}
\mathcal{H}^{M}_{\vec{k}}  = 
\begin{bmatrix}
\epsilon_{M} + \frac{k^2}{2 m_{M}} +  b  k_x k_y & 0\\
0 & \epsilon_{M} + \frac{k^2}{2 m_{M}} -  b  k_x k_y 
\end{bmatrix}
\end{align}
for electrons. The parameters entering Eqs.~\eqref{HGamma} and \eqref{HM} obtained from the fits to the tight-binding calculations~\cite{Kuroki2008,Cvetkovic2009} are tabulated in Ref.~(\onlinecite{Cvetkovic2013}). Below we set $a=c$, which corresponds to circular hole FSs. In this work we neglect the spin-orbit coupling, and at $\Gamma$, $\vec{k}=0$, the two Bloch states are degenerate. Equation \eqref{HM} neglects the admixture of $d_{XY}$ orbitals, and the parameter $b$ is the pocket ellipticity.

The Hamiltonians \eqref{HGamma} and \eqref{HM} describe the band structure shown in Fig.~\ref{fig:polarization}(b).
The band structure obtained by diagonalization of these Hamiltonians contains two hole pockets at $\Gamma$ with orbital content alternating between $d_{Xz}$ and $d_{Yz}$ with $\pi$ periodicity and two electron pockets at $M$. The electron pockets cross, and their outer parts contain an admixture of the $d_{XY}$ orbital. Here we neglect such an admixture while preserving the overall symmetry of the Hamiltonian.

\subsection{Raman coupling}

Raman scattering is a two-photon process. Its amplitude contains one part which is second order in the dipolar interaction and the first order in the coupling via the effective mass tensor. Assuming that the base frequency is detuned off the dipole transitions, it is customary to ignore the dipolar coupling. Under these circumstances the inelastic Raman scattering cross section as a function of the Raman shift $\omega$ is proportional to the imaginary part of the retarded Raman susceptibility $[\kappa^R(\vec{q},\omega)]''$.
We compute it from the corresponding Matsubara correlation function of the Raman vertices, 
\begin{align}\label{kappa_define}
\kappa(q) = 
\langle \hat{r}\hat{r} \rangle_{q} \, ,
\end{align} 
where the vector $q = (\vec{q},i\omega_m)$ includes the spatial wave vector $\vec{q}$, and Matsubara frequency $\omega_m$, and we denote $ \langle \hat{A} \hat{B} \rangle_{q} = \int_0^{T^{-1}} d \tau \exp(i \omega_m \tau) \langle A_{\vec{q}}(\tau) B_{-\vec{q}}(0) \rangle$. The experimental situation corresponds to $\vec{q}=0$ in Eq.~\eqref{kappa_define}, and the Raman susceptibility $\kappa^R(0,\omega)$ is obtained from $\kappa(q)$ by setting $\vec{q}=0$ and performing the analytical continuation, $i \omega_m \rightarrow \omega$. Below in writing the Matsubara frequency $\omega_m$, we omit the subscript $m$ for brevity.

The expression for the Raman vertices 
\begin{align}\label{r_vertex1}
\hat{r} = \sum_{\vec{k},\alpha}\sum_{i,j=1,2} 
c_{i,\vec{k}\alpha}^{\dag}  r^{\Gamma}_{i,j} c_{j,\vec{k}\alpha}
+
f_{i,\vec{k}\alpha}^{\dag}  r^{M}_{i,j} f_{j,\vec{k}\alpha} 
\end{align}
is fixed by the Hamiltonian, as formulated by Eqs.~\eqref{Hfree}, \eqref{HGamma}, and \eqref{HM}, as well as the polarization vectors of incoming and scattered photons, $\vec{e}^I$ and $\vec{e}^S$,
\begin{align}\label{r_vertex2}
r^{\Gamma(M)}_{i,j} = \sum_{\lambda\lambda'} e^I_{\lambda} e^S_{\lambda'} \frac{ \partial^2 \mathcal{H}_{ij}^{\Gamma(M)} }{\partial k_{\lambda} \partial k_{\lambda'}}\, .
\end{align}

In this work we focus on the $B_{2g}$ Raman configuration such that polarization vectors of incoming and scattered photons are $\vec{e}^I = (\hat{X} + \hat{Y})/ \sqrt{2} = \hat{x}$ and  $\vec{e}^S = (\hat{Y} - \hat{X})/ \sqrt{2}=\hat{y}$, respectively (see Fig.~\ref{fig:polarization}). The reason for this is twofold. 
First, the buildup of the low-energy $B_{2g}$ Raman intensity upon cooling is the dominant feature observed experimentally above $T_s$.~\cite{Gallais2013,Zhang2014,Thorsmolle2014}
Second, both orbital and the nematic fluctuations have $B_{2g}$ symmetry.
Indeed, Eq.~\eqref{r_vertex2} in combination with Eqs.~\eqref{HGamma} and \eqref{HM} gives
\begin{align}\label{r_vertex3}
r^{\Gamma} = a
\begin{bmatrix}
1 & 0 \\
0 & -1
\end{bmatrix}\, ,\quad
r^{M} = b
\begin{bmatrix}
1 & 0 \\
0 & -1
\end{bmatrix}\, .
\end{align}
Equation \eqref{r_vertex3} shows that photons in the $B_{2g}$ Raman configuration couple directly to orbital fluctuations.
 
\section{Effective action and Raman susceptibility}
 
We compute the Raman susceptibility as given by Eq.~\eqref{kappa_define} with the Raman vertex specified by Eqs.~\eqref{r_vertex1} and \eqref{r_vertex3}. In doing so we follow closely the derivation of Ref.~(\onlinecite{Fernandes2012}). To compute the Raman susceptibility we add to the quantum action the source term,
\begin{align}\label{source}
S_{J} = J_{\omega} \hat{r}_{-\omega} + J_{-\omega} \hat{r}_{\omega}
\end{align}
and the Raman susceptibility [Eq.~\eqref{kappa_define}] is obtained by a functional derivative of a free-energy functional, 
\begin{align}\label{source_1}
\kappa (\omega) = \frac{ \delta^2 \mathcal{F}[J] }{ \delta J_{\omega} \delta J_{- \omega}}, 
\end{align}
computed at $J_{\omega} = J_{-\omega} = 0$.

Here we focus on the spin interactions for definiteness and comment on the role of the orbital fluctuations.
In the purely magnetic scenario of the nematic transition we write the interaction in the form
\begin{align}\label{H_int}
\mathcal{H}_{int} = -\frac{1}{2} u_{s} \sum_{\vec{q}}\sum_{i=1,2} \vec{s}_{i,\vec{q}} \vec{s}_{i, - \vec{q}}\, ,
\end{align}
where the spin operator is diagonal in orbital index $i$,
\begin{align}\label{s}
\vec{s}_{i,\vec{q}} = \sum_{\vec{k}}\sum_{i=1,2}  c_{\vec{k}+\vec{q},i\alpha}^{\dag}
\vec{\sigma}_{\alpha\beta} f_{\vec{k},i\beta}\, ,
\end{align}
where $\vec{\sigma}_{\alpha\beta}$ are the Pauli matrices. 
The standard Hubbard-Stratonovich transformation amounts to the decoupling of the interaction term [Eq.~\eqref{H_int}] via the stripe magnetizations, $\vec{\Delta}^{X(Y)} \propto \sum_{\vec{k}} c_{\vec{k}+\vec{q}, 1(2)\alpha}^{\dag}
\vec{\sigma}_{\alpha\beta} f_{\vec{k},1(2)\beta}$.
The integration over fermion operators results in an effective action that closely resembles that of Ref.~(\onlinecite{Fernandes2012}),
\begin{align}\label{action_1}
S & [\vec{\Delta}^{X,Y},J_{\pm\omega}] =  \int_{q'} \chi^{-1}_{q'} \left( |\bm{\Delta}^X_{q'}|^2 + |\bm{\Delta}^Y_{q'}|^2 \right)
\notag \\
-&\frac{g}{2} 
\left[ |\Xi_{XY}(0)|^2 + |\Xi_{XY}(\omega)|^2 \right]+ 
\left[\lambda_{AL} J_{\omega}  \Xi_{XY}(\omega) + c.c.\right]\, ,
\end{align}
where we introduced the notation   
\begin{align}\label{Xi}
\Xi_{XY}(\omega)= \sum_{\Omega,\vec{q}}  
\left[\vec{\Delta}^X_{\vec{q},\omega + \Omega} \vec{\Delta}^X_{-\vec{q},-\Omega}
-\vec{\Delta}^Y_{\vec{q},\omega + \Omega} \vec{\Delta}^Y_{-\vec{q},-\Omega}\right].
\end{align}
For $\omega=0$, Eq.~\eqref{Xi} describes the classical contribution of the nematic fluctuations to the Ginzburg-Landau free energy, while $\Xi_{XY}(\omega)$ describes the quantum nematic fluctuation driven by the external source at the same frequency.
 
We now comment on Eq.~\eqref{action_1}. First, we omitted the term $ \propto [(\bm{\Delta}^X)^2 + (\bm{\Delta}^Y)^2]^2$ responsible for the renormalization of the spin susceptibility $\chi_q$ that iscrucially important for the nature of the magnetic and structural phase transition. Here we are not concerned with either the feedback of nematic fluctuations on magnetism or mapping out the phase diagram, and for that reason we do not include this term in the action keeping the spin susceptibility unrenormalized.

Second, the last term in Eq.~\eqref{action_1} describes the coupling of the Raman vertex to the spin-nematic order parameter $\propto (\bm{\Delta}^X)^2 - (\bm{\Delta}^Y)^2$ via the triangular Aslamazov-Larkin-like vertex $\lambda_{AL}$ evaluated in the Appendix \ref{app:A}. We have shown this vertex can be approximated by a frequency- and momentum-independent function. Most relevant for the present analysis is the weak temperature dependence of $\lambda_{AL}$ for temperatures not exceeding the mismatch between the electron and hole Fermi surfaces. Because the latter can be a few tens of meV, the above statements hold for most of the relevant temperatures. 

We further perform the second Hubbard-Stratonovich transformation by introducing the nematic field $\phi$.
We introduce the static nematic field $\phi_0$ and the quantum time-dependent nematic field $\phi(\tau) = \phi_{\omega} e^{ i \omega \tau}$, which corresponding to the two terms quartic in $\bm{\Delta}^{X(Y)}$ in Eq.~\eqref{action_1} . The resulting quadratic action reads
\begin{align}\label{action_2}
S =&   \sum_q (\chi_q^{-1} + \phi_0) | \bm{\Delta}^X_q|^2 + (\chi_q^{-1} - \phi_0)|\bm{\Delta}^Y_{q}|^2  
\notag \\
&+ 
\left[ (\phi_{-\omega} + \lambda_{AL}  J_{- \omega}) \Xi_{XY}(\omega)
 +c.c. \right]
\notag \\ 
&+
\frac{\phi_0^2}{ 2 g } + \frac{\left| \phi_{\omega} \right|^2 }{  g }\, .
\end{align}
The action \eqref{action_2} is quadratic with respect to the stripe order parameters, which thus can be integrated out explicitly. This procedure results in the effective action
\begin{align}\label{action_3}
  \exp(-\mathcal{F})&=\int d\bm{\Delta}^Xd\bm{\Delta}^Y \exp(-S)\notag\\
  =
  \exp&\left(-\frac{\phi_0^2}{ 2 g } - \frac{\left| \phi_{\omega} \right|^2 }{  g }  \right)\mathrm{Det}(\hat{\mathcal{D}}^3_+)\mathrm{Det}(\hat{\mathcal{D}}^3_-),
  \end{align}  
  where we have defined matrices in the Fourier space 
  \begin{equation}\label{action_4}
  \hat{\mathcal{D}}_{\pm}=\hat{\chi}^{-1}_{\pm}   \pm(\phi_{-\omega} + \lambda_{sc} J_{- \omega}) \hat{M}_+ \pm (\phi_{\omega} + \lambda_{sc} J_{ \omega}) \hat{M}_- ,
  \end{equation}
with $\hat{\chi}^{-1}_{\pm}  = \delta_{\vec{q}_1,\vec{q}_2} \delta_{\omega_1,\omega_2} (\chi^{-1}_{q_1} \pm \phi_0)$ and 
$\hat{M}_{\pm}= \delta_{\vec{q}_1,\vec{q}_2} \delta_{\omega_1, \omega_2 \pm \omega}$. With the help of the standard formula we convert the determinant into the trace of the logarithm, $\mathrm{Det}\hat{\mathcal{D}}=\exp(\mathrm{Tr}\ln\hat{\mathcal{D}})$,  and expand the resulting effective action up to second order in the nematic fields  and the source strength. Such an expansion amounts to the mean-field approximation that can be justified in the large-$N$ limit. In Ref.~(\onlinecite{Fernandes2012}) the thermodynamic properties of the same model were shown to be reasonably well captured in this approximation for $N=3$, and we employ it here as well and write
\begin{align}\label{action_5}
\mathcal{F}&=\frac{\phi_0^2}{ 2 g } + \frac{\left| \phi_{\omega} \right|^2}{g}
 -3 \phi_{0}^2  \Upsilon(0) 
-3 \left|(\phi_{\omega} + \lambda_{AL} J_{ \omega})\right|^2   \Upsilon(\omega) \, ,     
\end{align}
where the dynamical spin-nematic susceptibility 
\begin{align}\label{bare}
\Upsilon(\omega)=\sum_{\bm{q},\Omega} \chi(\bm{q},\Omega) \chi(\bm{q},\omega + \Omega)
\end{align}
has been introduced.

\begin{figure}[t!]
  \includegraphics[width=9cm]{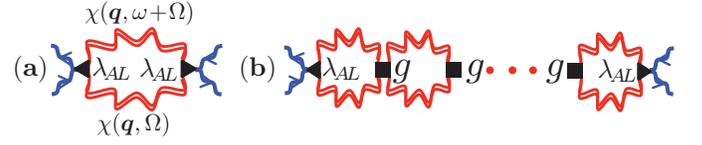}\\
  \caption{Feynman graphs illustrating results \eqref{bare} and \eqref{Raman-k}. The pair of thick blue wavy lines at the left and right ends of both graphs represents incoming and scattered photons. 
The black triangles represent the Aslamazov-Larkin triangular vertex giving rise to the coupling constant $\lambda_{AL}$ of light to the spin-nematic order parameter, $(\bm{\Delta}^X)^2-(\bm{\Delta}^Y)^2$, computed in the Appendix \ref{app:A}. The double wavy red lines denote the spin susceptibilities $\chi(\bm{q},\Omega)$. While the graph in (a) is not specific to the $XY$ geometry, the attraction in the nematic channel $g > 0$ makes it necessary to include the ladder diagrams shown in (b), giving rise to the quasielastic peak in $B_{2g}$ geometry.}\label{Fig-Worm}
\end{figure}

The action \eqref{action_5} is quadratic, and the functional derivative in Eq.~\eqref{source_1} gives, for the Raman susceptibility,
 \begin{align}\label{Raman-k}
\kappa(\omega)=3|\lambda_{AL}|^2\frac{\Upsilon(\omega)}{1-3g\Upsilon(\omega)};
\end{align}
see Fig.~\ref{Fig-Worm} for diagrammatic representation. To compute $\Upsilon(\omega)$ we use the standard finite-temperature Matsubara summation technique over the discrete frequencies followed by an analytic continuation to the real axis, $i \omega_n \rightarrow \omega + i 0$, to obtain the retarded spin-nematic correlation function $\Upsilon(i\omega_n)\to \Upsilon^R(\omega)$.

To this end, we evaluate the bare susceptibility (\ref{bare}) by converting the Matsubara sum over $\Omega$ into the complex integral 
\begin{align}
\Upsilon(i \omega_n) = \sum_{\bm{q}}
\oint \frac{ d z }{ 4 \pi i } \coth\frac{ z }{ 2 T } \chi(- i z +\omega_m,\bm{q}) \chi(-i z,\bm{q})  \, .
\end{align} 
The integrand has two branch cuts at $\Im(  z + i \omega_n) = 0$ and $ \Im( z) = 0$, where the product of two $\chi$ functions has breaks of analyticity. As a result of the analytic continuation process, we get
\begin{align}\label{sus}
&\Upsilon(\omega) = \sum_{\bm{q}}
\int \frac{d \Omega }{ 2 \pi } \coth \frac{ \Omega }{ 2 T } \notag\\ 
&\left[
\chi^R(\bm{q},\Omega + \omega) \Im \chi^R(\bm{q},\Omega) 
+
\Im \chi^R(\bm{q},\Omega) \chi^A(\bm{q},\Omega - \omega)
\right]
\end{align}   
To make further progress we use the standard expression for the spin-correlation function~\cite{Li2009,Tucker2012},
\begin{equation}\label{MMP}
\chi(\bm{q},\Omega_m)=\frac{c}{\xi^{-2}+(\bm{Q}-\bm{q})^2+|\Omega_m|/\gamma},
\end{equation}
where the important scale is
\begin{align}\label{tau_s}
\frac{1}{\tau_s} =\frac{ \gamma}{ \xi^{2} }\, ,
\end{align}
$c$ is a constant, $\gamma$ is the Landau damping coefficient, and $\bm{Q}=(\pi,\pi)$. 
Separating the real and imaginary components, one finds
\begin{align}\label{MMP-ret1}
\Im \chi^R(\bm{q}+\bm{Q},\Omega) = \chi'' =
c \xi^2  \frac{  \Omega \tau_s}{ (1+ q^2 \xi^{2} )^2 + \Omega^2 \tau_s^2  }, \end{align}
\begin{align}\label{MMP-ret2}
\Re \chi^R(\bm{q}+\bm{Q},\Omega) = \chi'=
c \xi^2 \frac{ 1 + q^2 \xi^2}{ (1+ q^2 \xi^{2} )^2 + \Omega^2 \tau_s^2  } .
\end{align}

The symmetry between the two stripelike spin-ordering arrangements is broken at the Ising-nematic-type transition. Here we assume the mean-field critical exponent $\nu = 1/2$, i.e.,  
\begin{align}\label{xi_define}
\xi(T) \approx l \sqrt{T_{N}/(T-T_N)},
\end{align}
with $T_{N}$ being the mean-field SDW transition temperature and $l$ being a microscopic length scale.
We emphasize that the mean-field transition temperature can be substantially lower than the observed SDW transition temperature, $T_N < T_{SDW}$. Equation~\eqref{MMP} also shows that the critical behavior of the static susceptibility $\chi(\Omega_m = 0,\bm{Q}) = c \xi^2 \propto (T - T_N)^{-1}$ is as prescribed by the mean field.

We proceed by substituting \eqref{MMP-ret1} and \eqref{MMP-ret2} into the general relation \eqref{sus} to get, for the imaginary part of the susceptibility,
\begin{align}\label{bare_5}
&\Upsilon''( \omega)=\gamma\int_0^{\infty} \frac{d x}{2 \pi} \int_{-\infty}^{\infty} \frac{d y}{ 2 \pi}  \frac{ y  \coth(y/2t)}{(1+x)^2 + y^2} \notag\\
\times&\left[\frac{ y+ w}{(1+x)^2 + (y + w)^2}-\frac{ y- w }{(1+x)^2 + (y - w)^2} 
\right].
\end{align}
Here we introduced dimensionless variables $x=(q\xi)^2$, $y = \Omega \tau_s$, $t = T \tau_s$, and $w = \omega \tau_s$. The above expression is general. In a view of Raman experiments, below we consider in detail the limiting case of high temperatures, $T\tau_s\gg1$, that corresponds to the regime of essentially classical fluctuations.  

\begin{figure}[t!]
  \includegraphics[width=9cm]{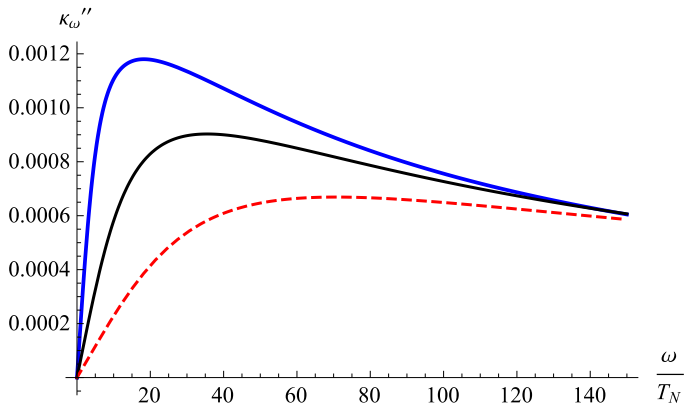}\\
  \caption{(color online) The modeling of the quasielastic peak in the Raman response function in accordance with Eq.~(\ref{QEP}), where $\kappa''$ is plotted in units of $3\lambda^2_{AL}\gamma$ for the following choice of fitting parameters from the bottom curve to the top one:
$\tau = 2$ (red dashed line), $\tau=1.5$ (black thin solid line), and $\tau = 1.25$ (blue solid thick line); $\bar{g}=0.2$, $\bar{\gamma}=5$, $L=10$. The peak grows when the structural transition is approached upon cooling.}\label{Fig-QEP-MMP}
\end{figure}

In the classical region when $t\gg1$, assuming not too high frequencies, $T>\omega$, one can approximate $\coth(y/2t)\approx 2t/y$. The double integral in Eq.~(\ref{bare_5}) can be then evaluated analytically,
\begin{equation}
\Upsilon''(\omega)=\frac{4T\gamma}{\omega}\ln\left[1+\left(\frac{\omega\xi^2}{2\gamma}\right)^2\right].
\end{equation}  
Since the integral decays at a scale $\sim 4 \tau_s$, the approximation made should be reasonable for all frequencies. Similarly, we evaluate the real part of the susceptibility,
\begin{align}\label{bare_8}
\Upsilon'( \omega)=&
2 \tau_s^{-1} \xi^2 \int_0^{\infty} \frac{d x}{4 \pi} \int_{-\infty}^{\infty} \frac{d y}{ 2 \pi}  \coth(y/2t) \notag
\\
& \times \frac{ y }{(1+x)^2 + y^2} 
\frac{ 1+ x }{(1+x)^2 + (y + w)^2} .
\end{align}
The integral in Eq.~\eqref{bare_8} is logarithmically divergent at ultraviolet.
We therefore isolate the divergence in Eq.~\eqref{bare_8} by focusing first on a static limit $\omega = 0$.
Then the difference is convergent and can be easily evaluated as
\begin{align}\label{bare_8aa}
\Upsilon'( \omega)= \Upsilon'(0)
-\frac{\gamma }{16 \pi } \arctan(\omega\tau_s/2) .
\end{align}
To evaluate the static susceptibility $\Upsilon'(0)$ we split the $y$-integration range in Eq.~\eqref{bare_8} into two regions, $y < t$ and $t< y< \Lambda \tau_s$, where the scale $\Lambda$ is the ultraviolet cutoff.
Making approximations $\coth(y/2t) \approx 2t/ y $ and  $\coth(y/2t) \approx \mathrm{sgn}(y) $ in the two respective intervals, the resulting integrals can be easily evaluated with the result
\begin{align}\label{bare_9}
\Upsilon'(0) \approx \frac{ T \xi^2 }{2 \pi^2} \arctan(T \tau_s) + \frac{ \tau_s^{-1} \xi^2 }{4 \pi^2} \ln\frac{\Lambda}{T},
\end{align}
where the arctangent can be further safely approximated by $\pi/2$. With these results at hand we find from Eq.~(\ref{Raman-k}) for the imaginary part of the Raman susceptibility  
\begin{equation}\label{QEP}
\kappa''(\omega)=3\lambda^2_{sc}\gamma\frac{I(\varpi,\tau)}{R^2(\varpi,\tau)+(4\pi^2\bar{g}\bar{\gamma})^2I^2(\varpi,\tau)}.
\end{equation}
Here we have introduced dimensionless frequency $\varpi=\omega/T_N$ and temperature $\tau=T/T_N$, and also two dimensionless functions, 
\begin{align}
&I(\varpi,\tau)=\frac{\tau}{\varpi}\ln\left[1+\left(\frac{\varpi}{2\pi\bar{\gamma}(\tau-1)}\right)^2\right],\\
&R(\varpi,\tau)=1-\bar{g}\left[\frac{\tau}{\tau-1}+\bar{\gamma}L-\frac{\pi\bar{\gamma}}{4}\arctan\left(\frac{\varpi}{2\pi\bar{\gamma}(\tau-1)}\right)\right],
\end{align}
where renormalized coupling $\bar{g}=3gT_Nl^2/4\pi$, decay rate $\bar{\gamma}=\gamma/\pi T_Nl^2$, and cutoff $L=\ln(\Lambda/T)$ should be used as fitting parameters; see Fig.~\ref{Fig-QEP-MMP} for results. 
The most prominent feature of our results is the critical enhancement of the Raman susceptibility upon approaching the structural transition with the characteristic buildup of the quasielastic scattering.

\section{Conclusions}

In this paper we investigated theoretically the low-energy Raman scattering in underdoped FeSCs.
The gross feature of the data is the quasielastic peak that gains in intensity and softens down at cooling above the structural transition. The phenomenon is observed exclusively in $B_{2g}$ Raman geometry. 
The Lorentzian-like frequency dependence of the $B_{2g}$ Raman susceptibility describes the relaxation dynamics with the relaxation rate given by the position of the maximum. The temperature dependence of the susceptibility indicates the freezing of the electronic $B_{2g}$ fluctuations at cooling. Such behavior is naturally associated with the tendency to long-range order which breaks the $B_{2g}$ symmetry. That is, the broad relaxation-like feature can be attributed to the critical slowdown associated with the approach to the discrete symmetry-breaking transition. Upon cooling, the system experiences locking in one of the two degenerate configurations related to the $C_4$ rotation for increasingly longer time intervals.

To understand the origin of the quasielastic peak as it appears in Eq.~\eqref{Raman-k}, note first that in the static limit the real part of the Raman susceptibility scales as $\sim (T - \theta)^{-1}$.
The temperature scale $\theta<T_s$, which can be explained in terms of the coupling of the electron nematic fluctuations and the orthorhombic lattice vibrations studied recently in Ref.~\onlinecite{Paul2014}.
These fluctuations add to the static nematic coupling constant,  $g_{st} = g + \bar{\gamma}^2/ C_s^0$, \cite{Gallais2015,Karahasanovic2015}.
Here $\bar{\gamma}$ is a nemato-elastic coupling constant and 
$C_s^0$ is the bare value of the orthorhombic elastic constant. 
The static coupling $g_{st}$ determines $T_s$.
Crucially, however, it is the dynamic rather than static nematic coupling constant $g<g_{st}$ that defines $\theta$ since the lattice response function has different static and dynamic limits \cite{Kontani2014}.
Correspondingly, the difference $T_s - \theta$ is expected to correlate with the reduction of $T_s$ in strain-controlled samples \cite{Chu2012}. This is indeed reported to be the case in the recent measurements \cite{Boehmer2015}. 

Distinguishing between different contributions to nematic correlations remains a challenge. Nevertheless, we can deduce the low-energy scattering by making the reasonable assumption on the  imaginary part of the bare response.  Assume that at low frequencies it scales as $\sim \omega/ \Gamma$, with $\Gamma$ being noncritical at $T = \theta$ and hence being a weakly temperature-dependent relaxation rate.
Then it follows from the denominator structure of  Eq.~\eqref{Raman-k} that at low frequencies $\kappa''(\omega) \sim ( T -\theta  + i \omega/\Gamma)^{-1}$
We thus see that the relaxation rate is suppressed by a factor of $T - \theta$ compared to the bare rate $\Gamma$. We conclude that the quasielastic scattering is the case of critical slowing down.

The intraband processes alone are insufficient to describe the large frequency width of the quasielastic scattering. Indeed, at zero momentum such transitions are forbidden, and the quasielastic peak is absent.~\cite{Platzman1965,Yamase2013} The small-momentum intraband transitions restricted to either $\Gamma$ or $M$ points are gapped and cannot account for quasielastic scattering either. The excitation of two electron-hole pairs at momentum close to the antiferromagnetic wave vector enables the relaxation of zero-momentum excitations by lifting the kinematical constraints. It was argued, however, that the phase-space limitations make the contributions of such processes to the relaxation rate scale as a cubic power of the frequency difference of scattered and incoming photons.~ \cite{Yamase2013} As our calculations demonstrate, this suppression is relevant only at very low frequencies, and for relevant temperatures and frequencies the scaling is essentially linear. In that regard this situation is very similar to that in cuprates.~\cite{Caprara2005} Even though cuprates are single-band rather than multiband materials, the processes that matter the most are confined to the vicinity of hot spots, i.e., the points on a FS connected by the antiferromagnetic wave vector.

\textit{Note added}. Recently, we became aware of arXiv:1503.07646 in which spin-nematic susceptibility was computed in a similar model. 

 \subsection*{Acknowledgments}

This work was inspired by discussions with G. Blumberg and benefited considerably from experimentally motivated insights shared with us by G. Blumberg, V.K. Thorsm\o lle and W.-L. Zhang.
We acknowledge the valuable discussions with E. Bettelheim,  
A. E. Boehmer, A.V. Chubukov, V. Cvetkovic, R.M. Fernandes, U. Karahasanovic, D. Orgad, J. Schmalian, and O. Vafek.
The work of M.K. was supported by the University of Iowa and Hebrew University of Jerusalem. 
The work of A.L. was supported by NSF Grant DMR-1401908.

\begin{appendix}
\section{Derivation of the effective action in terms of the stripe magnetizations}
\label{app:A}
\begin{widetext}
The calculations in this appendix are an extension of the corresponding derivations in Re.~(\onlinecite{Fernandes2012}).
We introduce the eight-component spinor
\begin{align}\label{app:1}
\Psi_{\vec{k}}^{\dag} =
\left(c_{1,\vec{k} \uparrow}^{\dag}, c_{2,\vec{k} \uparrow}^{\dag},f_{1,\vec{k} \uparrow}^{\dag},f_{2,\vec{k} \uparrow}^{\dag}; c_{1,\vec{k} \downarrow}^{\dag}, c_{2,\vec{k} \downarrow}^{\dag}, f_{1,\vec{k} \downarrow}^{\dag}, f_{2,\vec{k} \downarrow}^{\dag} \right)
\end{align}
in the direct product of orbital and spin spaces.
Upon the introduction of the stripe magnetizations, $\vec{\Delta}^{X,Y}$ via the Hubbard-Stratonovich transformation the action takes the form
\begin{align}\label{app:2}
S[\Psi,\vec{\Delta}^{X},\vec{\Delta}^{Y}, J_{\omega}, J_{- \omega}] =
-\int_k \Psi^{\dag}_{k} \mathcal{G}^{-1}_{k,k'} \Psi_{k'} + \frac{2}{u_s} \int_q
\left[ \left|\bm{\Delta}^X(q)\right|^2 + \left|\bm{\Delta}^Y(q)\right|^2\right]\, , 
\end{align}
where we denote $k = (\vec{k},i \epsilon)$, and the Green's function is
\begin{align}\label{app:3}
  \mathcal{G}^{-1} =  \mathcal{G}^{-1}_{0} - \mathcal{V}\, ,
\end{align}
where
\begin{align}\label{app:4}  
\mathcal{V}=   \mathcal{V}^{\Delta} - \mathcal{V}^{J_{\omega}}- \mathcal{V}^{J_{-\omega}}\, .
\end{align}
The free Green's function in Eq.~\eqref{app:3} is
\begin{align}\label{app:5}
\mathcal{G}_{0;k,k'} =
\delta_{k,k'} 
\begin{bmatrix}
\mathcal{G}^{\Gamma}_{\vec{k},\epsilon} & 0 \\
0 & \mathcal{G}^{M}_{\vec{k},\epsilon}
\end{bmatrix} \otimes \mathbb{I} \, ,
\end{align}
where the two hole and electron Green's functions are two-dimensional matrices expressed in a standard way,
\begin{align}\label{app:5a}
\mathcal{G}^{\Gamma,M}_{\vec{k},\epsilon} = 
\left(i \epsilon - \mathcal{H}^{\Gamma,M}_{\vec{k}} + E_F \right)^{-1}
\end{align}
through the hole and electron Hamiltonians \eqref{HGamma} and \eqref{HM} with energies counted relative to the Fermi level. We further have
\begin{align}\label{app:6}
\mathcal{V}^{\Delta}_{k,k'} = 
-\int_q 
\delta_{k+q,k'}
\begin{bmatrix}
0 & 0 &  \vec{\Delta}^X(q) & 0 \\
0 & 0 & 0 &  \vec{\Delta}^Y(q) \\
 \vec{\Delta}^X(q) & 0 & 0 & 0 \\
0 &  \vec{\Delta}^Y(q) & 0 & 0 
\end{bmatrix}
\otimes \vec{\sigma} \, .
\end{align}
The source term according to Eqs.~\eqref{r_vertex1} and \eqref{source}  reads,
\begin{align}\label{app:7}
\mathcal{V}^{J_{\pm \omega}}_{k,k'} = 
J_{\pm \omega} \delta_{\vec{k},\vec{k}'}
\delta_{\epsilon \pm \omega,\epsilon}
\begin{bmatrix}
r^{\Gamma} & 0  \\
0 & r^{M}  
\end{bmatrix}
\otimes \mathbb{I}
\end{align}
where the two-dimensional matrices $r^{\Gamma,M}$ are defined by Eq.~\eqref{r_vertex3}.
In the presence of the source \eqref{app:7}, it is necessary to keep the term of the third order in $\mathcal{V}$ in the expansion of the free energy,
\begin{align}\label{app:8}
S&[\vec{\Delta}^X,\vec{\Delta}^Y,J_{\omega},J_{-\omega}]
=
\frac{1}{2} \mathrm{Tr}(\mathcal{G}_0 \mathcal{V})^2+
\frac{1}{3} \mathrm{Tr}(\mathcal{G}_0 \mathcal{V})^3
+
\frac{1}{4} \mathrm{Tr}(\mathcal{G}_0 \mathcal{V})^4  
+ \frac{2}{u_s} \int_q
\left[\bm{\Delta}^X(q)|^2 + |\bm{\Delta}^Y(q)|^2\right],
\end{align}
which is of prime interest for us in this work. While even-order terms in Eq.~\eqref{app:8} were considered in detail in Ref.~(\onlinecite{Fernandes2012}), here we focus on the third-order term of Eq.~\eqref{app:8}.
We specifically determine the contribution to the effective action that is  linear in the sources $J_{\pm \omega}$ and quadratic in $\vec{\Delta}^{X,Y}$.
Such terms have the form
\begin{align}\label{app:10}
\mathrm{Tr} & (\mathcal{V}^{J_{\pm\omega}} \mathcal{G}_0\mathcal{V}^{\Delta}\mathcal{G}_0 \mathcal{V}^{\Delta}\mathcal{G}_0 )
= 
a  J_{\pm \omega} \sum_{\ell} \mathrm{Tr} \left\{
\begin{bmatrix}
1 & 0  \\
0 & -1 
\end{bmatrix}
\mathcal{G}^{\Gamma}_{\vec{k}+\vec{q},\epsilon\pm\omega} 
\begin{bmatrix}
\left( \bm{\Delta}^X_{\vec{q},\Omega \pm\omega}\right)_{\ell} & 0  \\
0 & \left( \bm{\Delta}^Y_{\vec{q},\Omega \pm \omega} \right)_{\ell}
\end{bmatrix}
 \mathcal{G}^{M}_{\vec{k},\epsilon-\Omega} 
\begin{bmatrix}
\left( \bm{\Delta}^X_{\vec{q},-\Omega}\right)_{\ell} & 0  \\
0 & \left(\bm{\Delta}^Y_{\vec{q},-\Omega} \right)_{\ell}
\end{bmatrix}
\mathcal{G}^{\Gamma}_{\vec{k},\epsilon} 
\right\}
\notag \\
& +
b \sum_{\ell} J_{\pm \omega} \mathrm{Tr} \left\{
\begin{bmatrix}
1 & 0  \\
0 & -1 
\end{bmatrix}
\mathcal{G}^{M}_{\vec{k}+\vec{q},\epsilon \pm \omega} 
\begin{bmatrix}
\left( \bm{\Delta}^X_{\vec{q},\Omega \pm\omega}\right)_{\ell} & 0  \\
0 & \left( \bm{\Delta}^Y_{\vec{q},\Omega \pm \omega} \right)_{\ell}
\end{bmatrix}
 \mathcal{G}^{\Gamma}_{\vec{k},\epsilon-\Omega} 
\begin{bmatrix}
\left( \bm{\Delta}^X_{\vec{q},-\Omega}\right)_{\ell} & 0  \\
0 & \left(\bm{\Delta}^Y_{\vec{q},-\Omega} \right)_{\ell}
\end{bmatrix}
\mathcal{G}^{M}_{\vec{k},\epsilon} 
\right\} + c.c.
\end{align}
In obtaining Eq.~\eqref{app:10} we used the explicit form [Eq.~\eqref{r_vertex3}] of the Raman vertices $r^{\Gamma,M}$ and the block-diagonal structure of all three matrices given by Eqs.~\eqref{app:5}, \eqref{app:6}, and \eqref{app:7}. Furthermore the trace operation over the spin indices results in a summation over the Cartesian coordinates of the vectors of the stripe magnetizations $\vec{\Delta}^{X,Y}$ labeled by the index $\ell = x,y,z$. The trace operation in Eq.~\eqref{app:10} includes, in addition to the usual trace of the two-dimensional matrices, the summation over fermion frequencies and momenta $(i \epsilon,\vec{k})$ and boson frequencies and momenta $(i \Omega,\vec{q})$.

The integral of Eq.~\eqref{app:10} over the fermion frequencies and momenta is convergent at the upper limit. As a result, assuming that the Fermi energy is much larger than the typical energy carried by the magnetic fluctuation, we can neglect the boson momenta and frequency, $\vec{q}$ and $\Omega$, in the arguments of fermion Green's functions. Furthermore, since the energy splitting of electronic bands is normally smaller than the Fermi energy, we can neglect it and approximate the hole Hamiltonian by a scalar function. In terms of the Green's function we have 
\begin{align}\label{app:11}
\mathcal{G}^{\Gamma}_{\vec{k},\epsilon} \approx 
\bar{\mathcal{G}}^{\Gamma}_{\vec{k},\epsilon} = (i \epsilon - \xi_h)^{-1}\, ,
\end{align}
where $\xi_h$ is the energy of the hole band relative to the Fermi energy with splitting neglected.
Similarly, by neglecting the ellipticity-related energy that is small on the scale of the Fermi energy we arrive at the scalar Green's function for electrons, 
\begin{align}\label{app:12}
\mathcal{G}^{M}_{\vec{k},\epsilon} \approx 
\bar{\mathcal{G}}^{M}_{\vec{k},\epsilon} = (i \epsilon - \xi_e)^{-1}\, ,
\end{align}
where $\xi_e$ is the energy of the electron bands relative to the Fermi energy with ellipticity neglected.
Clearly, the above approximations make the corresponding Green's function denoted by $\bar{\mathcal{G}}^{\Gamma}_{\vec{k},\epsilon}$ and $\bar{\mathcal{G}}^{M}_{\vec{k},\epsilon}$ scalar and allow us to rewrite Eq.~\eqref{app:11} as
\begin{align}\label{app:13}
&\mathrm{Tr} (\mathcal{V}^{J_{\pm \omega}} \mathcal{G}_0\mathcal{V}^{\Delta}\mathcal{G}_0 \mathcal{V}^{\Delta}\mathcal{G}_0 )
\approx
J_{\pm \omega}
\Xi_{XY}(\pm \omega) 
\int_{k} 
\left[ a \left(\bar{\mathcal{G}}^{\Gamma}_{\vec{k},\epsilon}\right)^2
\bar{\mathcal{G}}^{M}_{\vec{k},\epsilon}
+
b \left(\bar{\mathcal{G}}^{M}_{\vec{k},\epsilon}\right)^2
\bar{\mathcal{G}}^{\Gamma}_{\vec{k},\epsilon}
\right]\, .
\end{align} 
Identifying the last term of Eq.~\eqref{action_1} with Eq.~\eqref{app:13}, we finally arrive at the expression for the triangular vertex,
\begin{align}\label{app:14}
\lambda_{AL} = \int_{k} 
\left[ a \left(\bar{\mathcal{G}}^{\Gamma}_{\vec{k},\epsilon}\right)^2
\bar{\mathcal{G}}^{M}_{\vec{k},\epsilon}
+
b \left(\bar{\mathcal{G}}^{M}_{\vec{k},\epsilon}\right)^2
\bar{\mathcal{G}}^{\Gamma}_{\vec{k},\epsilon}
\right]\, ,
\end{align}
where the expression for $\Xi_{XY}(\omega)$ is given in Eq.~\eqref{Xi}. For definiteness, we evaluate the first term $\propto a$ in expression \eqref{app:14}. For simplicity we assume $\xi_h = -\xi_e + \delta_{he} = \xi$ and take the density of states to be a constant $\nu_0$ for both electrons and holes. Under these assumptions the substitution of the Green's functions in Eqs.~\eqref{app:11} and \eqref{app:12} in Eq.~\eqref{app:14}, followed by the integration over $\xi$, yields
\begin{align}\label{app:15}
 \int_{k} 
  \left(\bar{\mathcal{G}}^{\Gamma}_{\vec{k},\epsilon}\right)^2
\bar{\mathcal{G}}^{M}_{\vec{k},\epsilon}
=
2 \pi i T \nu_0  \sum_{\epsilon_n >0} 
\left[\left(\frac{1}{ 2 i \epsilon_n + \delta_{eh} }\right)^2 - 
\left(\frac{1}{ 2 i \epsilon_n - \delta_{eh} }\right)^2 \right].
\end{align}
As the Matsubara frequencies are of the form $\epsilon_n = 2 \pi (n+1/2)T$, we obtain from Eq.~\eqref{app:15}
\begin{align}\label{app:16}
 \int_{k} 
  \left(\bar{\mathcal{G}}^{\Gamma}_{\vec{k},\epsilon}\right)^2
\bar{\mathcal{G}}^{M}_{\vec{k},\epsilon}
=
\frac{\nu_0}{ \pi T} \mathrm{Im}\left[ \psi'\left( \frac{1}{2} - i \frac{ \delta_{eh}}{2 \pi T} \right)\right]\, ,
\end{align}
where $\psi'(x)$ is the derivative of the digamma function.
An expression similar to Eq.~\eqref{app:16} holds for the contribution of the coupling of light to electrons, and we obtain
\begin{align}\label{app:17}
\lambda_{AL} = (a+b)
\frac{\nu_0}{ \pi T} \mathrm{Im}\left[ \psi'\left( \frac{1}{2} - i \frac{ \delta_{eh}}{2 \pi T} \right)\right]\, .
\end{align}
We note that the contributions of electrons and holes are additive.
As $\delta_{eh}$ is typically a few tens of meV, we expect the inequality $T \lesssim \delta_{eh}$ to hold.
As the asymptotic expansion $\mathrm{Im}[\psi'(1/2 - i x)] \approx 1/x$ holds already for $x \gtrsim 0.25$, we can write for the Aslamazov-Larkin vertex
\begin{align}\label{app:18}
\lambda_{AL} \approx (a+b)
\frac{2\nu_0}{ \delta_{eh}} \, .
\end{align}
We conclude that $\lambda_{AL}$ is insensitive to the temperature variation in the relevant temperature range and is suppressed with increasing mismatch between the hole and electron Fermi surfaces.
This result is in agreement with the alternative calculation for a different model. \cite{Kontani2014}
\end{widetext}
\end{appendix}


\begin{thebibliography}{49}
\expandafter\ifx\csname natexlab\endcsname\relax\def\natexlab#1{#1}\fi
\expandafter\ifx\csname bibnamefont\endcsname\relax
  \def\bibnamefont#1{#1}\fi
\expandafter\ifx\csname bibfnamefont\endcsname\relax
  \def\bibfnamefont#1{#1}\fi
\expandafter\ifx\csname citenamefont\endcsname\relax
  \def\citenamefont#1{#1}\fi
\expandafter\ifx\csname url\endcsname\relax
  \def\url#1{\texttt{#1}}\fi
\expandafter\ifx\csname urlprefix\endcsname\relax\def\urlprefix{URL }\fi
\providecommand{\bibinfo}[2]{#2}
\providecommand{\eprint}[2][]{\url{#2}}

\bibitem[{\citenamefont{Mazin}(2010)}]{Mazin2010}
\bibinfo{author}{\bibfnamefont{I.~I.} \bibnamefont{Mazin}},
  \bibinfo{journal}{Nature} \textbf{\bibinfo{volume}{464}},
  \bibinfo{pages}{183} (\bibinfo{year}{2010}).

\bibitem[{\citenamefont{Paglione and Greene}(2010)}]{Paglione2010}
\bibinfo{author}{\bibfnamefont{J.}~\bibnamefont{Paglione}} \bibnamefont{and}
  \bibinfo{author}{\bibfnamefont{R.~L.} \bibnamefont{Greene}},
  \bibinfo{journal}{Nat Phys} \textbf{\bibinfo{volume}{6}},
  \bibinfo{pages}{645} (\bibinfo{year}{2010}).

\bibitem[{\citenamefont{Johnston}(2010)}]{Johnston2010}
\bibinfo{author}{\bibfnamefont{D.~C.} \bibnamefont{Johnston}},
  \bibinfo{journal}{Advances in Physics} \textbf{\bibinfo{volume}{59}},
  \bibinfo{pages}{803} (\bibinfo{year}{2010}).

\bibitem[{\citenamefont{Stewart}(2011)}]{Stewart2011}
\bibinfo{author}{\bibfnamefont{G.~R.} \bibnamefont{Stewart}},
  \bibinfo{journal}{Rev. Mod. Phys.} \textbf{\bibinfo{volume}{83}},
  \bibinfo{pages}{1589} (\bibinfo{year}{2011}).

\bibitem[{\citenamefont{Basov and Chubukov}(2011)}]{Basov2011}
\bibinfo{author}{\bibfnamefont{D.~N.} \bibnamefont{Basov}} \bibnamefont{and}
  \bibinfo{author}{\bibfnamefont{A.~V.} \bibnamefont{Chubukov}},
  \bibinfo{journal}{Nat Phys} \textbf{\bibinfo{volume}{7}},
  \bibinfo{pages}{272} (\bibinfo{year}{2011}).

\bibitem[{\citenamefont{Kamihara et~al.}(2006)\citenamefont{Kamihara,
  Hiramatsu, Hirano, Kawamura, Yanagi, Kamiya, and Hosono}}]{Kamihara2006}
\bibinfo{author}{\bibfnamefont{Y.}~\bibnamefont{Kamihara}},
  \bibinfo{author}{\bibfnamefont{H.}~\bibnamefont{Hiramatsu}},
  \bibinfo{author}{\bibfnamefont{M.}~\bibnamefont{Hirano}},
  \bibinfo{author}{\bibfnamefont{R.}~\bibnamefont{Kawamura}},
  \bibinfo{author}{\bibfnamefont{H.}~\bibnamefont{Yanagi}},
  \bibinfo{author}{\bibfnamefont{T.}~\bibnamefont{Kamiya}}, \bibnamefont{and}
  \bibinfo{author}{\bibfnamefont{H.}~\bibnamefont{Hosono}},
  \bibinfo{journal}{Journal of the American Chemical Society}
  \textbf{\bibinfo{volume}{128}}, \bibinfo{pages}{10012}
  (\bibinfo{year}{2006}), \bibinfo{note}{pMID: 16881620},
  \eprint{http://dx.doi.org/10.1021/ja063355c}.

\bibitem[{\citenamefont{Kamihara et~al.}(2008)\citenamefont{Kamihara, Watanabe,
  Hirano, and Hosono}}]{Kamihara2008}
\bibinfo{author}{\bibfnamefont{Y.}~\bibnamefont{Kamihara}},
  \bibinfo{author}{\bibfnamefont{T.}~\bibnamefont{Watanabe}},
  \bibinfo{author}{\bibfnamefont{M.}~\bibnamefont{Hirano}}, \bibnamefont{and}
  \bibinfo{author}{\bibfnamefont{H.}~\bibnamefont{Hosono}},
  \bibinfo{journal}{J. Am. Chem. Soc.} \textbf{\bibinfo{volume}{130}},
  \bibinfo{pages}{3296} (\bibinfo{year}{2008}).

\bibitem[{\citenamefont{Chen et~al.}(2008)\citenamefont{Chen, Wu, Wu, Liu,
  Chen, and Fang}}]{Chen2008}
\bibinfo{author}{\bibfnamefont{X.~H.} \bibnamefont{Chen}},
  \bibinfo{author}{\bibfnamefont{T.}~\bibnamefont{Wu}},
  \bibinfo{author}{\bibfnamefont{G.}~\bibnamefont{Wu}},
  \bibinfo{author}{\bibfnamefont{R.~H.} \bibnamefont{Liu}},
  \bibinfo{author}{\bibfnamefont{H.}~\bibnamefont{Chen}}, \bibnamefont{and}
  \bibinfo{author}{\bibfnamefont{D.~F.} \bibnamefont{Fang}},
  \bibinfo{journal}{Nature} \textbf{\bibinfo{volume}{453}},
  \bibinfo{pages}{761} (\bibinfo{year}{2008}).

\bibitem[{\citenamefont{Sefat et~al.}(2008)\citenamefont{Sefat, Jin, McGuire,
  Sales, Singh, and Mandrus}}]{Sefat2008}
\bibinfo{author}{\bibfnamefont{A.~S.} \bibnamefont{Sefat}},
  \bibinfo{author}{\bibfnamefont{R.}~\bibnamefont{Jin}},
  \bibinfo{author}{\bibfnamefont{M.~A.} \bibnamefont{McGuire}},
  \bibinfo{author}{\bibfnamefont{B.~C.} \bibnamefont{Sales}},
  \bibinfo{author}{\bibfnamefont{D.~J.} \bibnamefont{Singh}}, \bibnamefont{and}
  \bibinfo{author}{\bibfnamefont{D.}~\bibnamefont{Mandrus}},
  \bibinfo{journal}{Phys. Rev. Lett.} \textbf{\bibinfo{volume}{101}},
  \bibinfo{pages}{117004} (\bibinfo{year}{2008}).

\bibitem[{\citenamefont{Chu et~al.}(2009)\citenamefont{Chu, Analytis,
  Kucharczyk, and Fisher}}]{Chu2009}
\bibinfo{author}{\bibfnamefont{J.-H.} \bibnamefont{Chu}},
  \bibinfo{author}{\bibfnamefont{J.~G.} \bibnamefont{Analytis}},
  \bibinfo{author}{\bibfnamefont{C.}~\bibnamefont{Kucharczyk}},
  \bibnamefont{and} \bibinfo{author}{\bibfnamefont{I.~R.}
  \bibnamefont{Fisher}}, \bibinfo{journal}{Phys. Rev. B}
  \textbf{\bibinfo{volume}{79}}, \bibinfo{pages}{014506}
  (\bibinfo{year}{2009}).

\bibitem[{\citenamefont{Takahashi et~al.}(2008)\citenamefont{Takahashi, Okada,
  Igawa, Arii, Kamihara, Matsuishi, Hirano, Hosono, Matsubayashi, and
  Uwatoko}}]{Takahashi2008}
\bibinfo{author}{\bibfnamefont{H.}~\bibnamefont{Takahashi}},
  \bibinfo{author}{\bibfnamefont{H.}~\bibnamefont{Okada}},
  \bibinfo{author}{\bibfnamefont{K.}~\bibnamefont{Igawa}},
  \bibinfo{author}{\bibfnamefont{K.}~\bibnamefont{Arii}},
  \bibinfo{author}{\bibfnamefont{Y.}~\bibnamefont{Kamihara}},
  \bibinfo{author}{\bibfnamefont{S.}~\bibnamefont{Matsuishi}},
  \bibinfo{author}{\bibfnamefont{M.}~\bibnamefont{Hirano}},
  \bibinfo{author}{\bibfnamefont{H.}~\bibnamefont{Hosono}},
  \bibinfo{author}{\bibfnamefont{K.}~\bibnamefont{Matsubayashi}},
  \bibnamefont{and} \bibinfo{author}{\bibfnamefont{Y.}~\bibnamefont{Uwatoko}},
  \bibinfo{journal}{Journal of the Physical Society of Japan}
  \textbf{\bibinfo{volume}{77}}, \bibinfo{pages}{78} (\bibinfo{year}{2008}),
  \eprint{http://dx.doi.org/10.1143/JPSJS.77SC.78}.

\bibitem[{\citenamefont{Hamlin et~al.}(2008)\citenamefont{Hamlin, Baumbach,
  Zocco, Sayles, and Maple}}]{Hamlin2008}
\bibinfo{author}{\bibfnamefont{J.~J.} \bibnamefont{Hamlin}},
  \bibinfo{author}{\bibfnamefont{R.~E.} \bibnamefont{Baumbach}},
  \bibinfo{author}{\bibfnamefont{D.~A.} \bibnamefont{Zocco}},
  \bibinfo{author}{\bibfnamefont{T.~A.} \bibnamefont{Sayles}},
  \bibnamefont{and} \bibinfo{author}{\bibfnamefont{M.~B.} \bibnamefont{Maple}},
  \bibinfo{journal}{Journal of Physics: Condensed Matter}
  \textbf{\bibinfo{volume}{20}}, \bibinfo{pages}{365220}
  (\bibinfo{year}{2008}).

\bibitem[{\citenamefont{Chubukov et~al.}(2008)\citenamefont{Chubukov, Efremov,
  and Eremin}}]{Chubukov2008}
\bibinfo{author}{\bibfnamefont{A.~V.} \bibnamefont{Chubukov}},
  \bibinfo{author}{\bibfnamefont{D.~V.} \bibnamefont{Efremov}},
  \bibnamefont{and} \bibinfo{author}{\bibfnamefont{I.}~\bibnamefont{Eremin}},
  \bibinfo{journal}{Phys. Rev. B} \textbf{\bibinfo{volume}{78}},
  \bibinfo{pages}{134512} (\bibinfo{year}{2008}).

\bibitem[{\citenamefont{Chubukov}(2009)}]{Chubukov2009b}
\bibinfo{author}{\bibfnamefont{A.}~\bibnamefont{Chubukov}},
  \bibinfo{journal}{Physica C: Superconductivity}
  \textbf{\bibinfo{volume}{469}}, \bibinfo{pages}{640 } (\bibinfo{year}{2009}),
  ISSN \bibinfo{issn}{0921-4534}, \bibinfo{note}{superconductivity in
  Iron-Pnictides}.

\bibitem[{\citenamefont{Chubukov}(2012)}]{Chubukov2012}
\bibinfo{author}{\bibfnamefont{A.}~\bibnamefont{Chubukov}},
  \bibinfo{journal}{Annual Review of Condensed Matter Physics}
  \textbf{\bibinfo{volume}{3}}, \bibinfo{pages}{57} (\bibinfo{year}{2012}).

\bibitem[{\citenamefont{Fernandes et~al.}(2012)\citenamefont{Fernandes,
  Chubukov, Knolle, Eremin, and Schmalian}}]{Fernandes2012}
\bibinfo{author}{\bibfnamefont{R.~M.} \bibnamefont{Fernandes}},
  \bibinfo{author}{\bibfnamefont{A.~V.} \bibnamefont{Chubukov}},
  \bibinfo{author}{\bibfnamefont{J.}~\bibnamefont{Knolle}},
  \bibinfo{author}{\bibfnamefont{I.}~\bibnamefont{Eremin}}, \bibnamefont{and}
  \bibinfo{author}{\bibfnamefont{J.}~\bibnamefont{Schmalian}},
  \bibinfo{journal}{Phys. Rev. B} \textbf{\bibinfo{volume}{85}},
  \bibinfo{pages}{024534} (\bibinfo{year}{2012}).

\bibitem[{\citenamefont{Chu et~al.}(2010)\citenamefont{Chu, Analytis, De~Greve,
  McMahon, Islam, Yamamoto, and Fisher}}]{Chu2010}
\bibinfo{author}{\bibfnamefont{J.-H.} \bibnamefont{Chu}},
  \bibinfo{author}{\bibfnamefont{J.~G.} \bibnamefont{Analytis}},
  \bibinfo{author}{\bibfnamefont{K.}~\bibnamefont{De~Greve}},
  \bibinfo{author}{\bibfnamefont{P.~L.} \bibnamefont{McMahon}},
  \bibinfo{author}{\bibfnamefont{Z.}~\bibnamefont{Islam}},
  \bibinfo{author}{\bibfnamefont{Y.}~\bibnamefont{Yamamoto}}, \bibnamefont{and}
  \bibinfo{author}{\bibfnamefont{I.~R.} \bibnamefont{Fisher}},
  \bibinfo{journal}{Science} \textbf{\bibinfo{volume}{329}},
  \bibinfo{pages}{824} (\bibinfo{year}{2010}),
  \eprint{http://www.sciencemag.org/content/329/5993/824.full.pdf}.

\bibitem[{\citenamefont{Chu et~al.}(2012)\citenamefont{Chu, Kuo, Analytis, and
  Fisher}}]{Chu2012}
\bibinfo{author}{\bibfnamefont{J.-H.} \bibnamefont{Chu}},
  \bibinfo{author}{\bibfnamefont{H.-H.} \bibnamefont{Kuo}},
  \bibinfo{author}{\bibfnamefont{J.~G.} \bibnamefont{Analytis}},
  \bibnamefont{and} \bibinfo{author}{\bibfnamefont{I.~R.}
  \bibnamefont{Fisher}}, \bibinfo{journal}{Science}
  \textbf{\bibinfo{volume}{337}}, \bibinfo{pages}{710} (\bibinfo{year}{2012}),
  \eprint{http://www.sciencemag.org/content/337/6095/710.full.pdf}.

\bibitem[{\citenamefont{Mirri et~al.}(2014)\citenamefont{Mirri, Dusza,
  Bastelberger, Chu, Kuo, Fisher, and Degiorgi}}]{Mirri2014}
\bibinfo{author}{\bibfnamefont{C.}~\bibnamefont{Mirri}},
  \bibinfo{author}{\bibfnamefont{A.}~\bibnamefont{Dusza}},
  \bibinfo{author}{\bibfnamefont{S.}~\bibnamefont{Bastelberger}},
  \bibinfo{author}{\bibfnamefont{J.-H.} \bibnamefont{Chu}},
  \bibinfo{author}{\bibfnamefont{H.-H.} \bibnamefont{Kuo}},
  \bibinfo{author}{\bibfnamefont{I.~R.} \bibnamefont{Fisher}},
  \bibnamefont{and} \bibinfo{author}{\bibfnamefont{L.}~\bibnamefont{Degiorgi}},
  \bibinfo{journal}{Phys. Rev. B} \textbf{\bibinfo{volume}{89}},
  \bibinfo{pages}{060501} (\bibinfo{year}{2014}).

\bibitem[{\citenamefont{Fernandes et~al.}(2014)\citenamefont{Fernandes,
  Chubukov, and Schmalian}}]{Fernandes2014}
\bibinfo{author}{\bibfnamefont{R.~M.} \bibnamefont{Fernandes}},
  \bibinfo{author}{\bibfnamefont{A.~V.} \bibnamefont{Chubukov}},
  \bibnamefont{and}
  \bibinfo{author}{\bibfnamefont{J.}~\bibnamefont{Schmalian}},
  \bibinfo{journal}{Nat Phys} \textbf{\bibinfo{volume}{10}},
  \bibinfo{pages}{97} (\bibinfo{year}{2014}).

\bibitem[{\citenamefont{Kr\"uger et~al.}(2009)\citenamefont{Kr\"uger, Kumar,
  Zaanen, and van~den Brink}}]{Kruger2009}
\bibinfo{author}{\bibfnamefont{F.}~\bibnamefont{Kr\"uger}},
  \bibinfo{author}{\bibfnamefont{S.}~\bibnamefont{Kumar}},
  \bibinfo{author}{\bibfnamefont{J.}~\bibnamefont{Zaanen}}, \bibnamefont{and}
  \bibinfo{author}{\bibfnamefont{J.}~\bibnamefont{van~den Brink}},
  \bibinfo{journal}{Phys. Rev. B} \textbf{\bibinfo{volume}{79}},
  \bibinfo{pages}{054504} (\bibinfo{year}{2009}).

\bibitem[{\citenamefont{Lv et~al.}(2009)\citenamefont{Lv, Wu, and
  Phillips}}]{Lv2009}
\bibinfo{author}{\bibfnamefont{W.}~\bibnamefont{Lv}},
  \bibinfo{author}{\bibfnamefont{J.}~\bibnamefont{Wu}}, \bibnamefont{and}
  \bibinfo{author}{\bibfnamefont{P.}~\bibnamefont{Phillips}},
  \bibinfo{journal}{Phys. Rev. B} \textbf{\bibinfo{volume}{80}},
  \bibinfo{pages}{224506} (\bibinfo{year}{2009}).

\bibitem[{\citenamefont{Lee et~al.}(2009)\citenamefont{Lee, Yin, and
  Ku}}]{Lee2009a}
\bibinfo{author}{\bibfnamefont{C.-C.} \bibnamefont{Lee}},
  \bibinfo{author}{\bibfnamefont{W.-G.} \bibnamefont{Yin}}, \bibnamefont{and}
  \bibinfo{author}{\bibfnamefont{W.}~\bibnamefont{Ku}}, \bibinfo{journal}{Phys.
  Rev. Lett.} \textbf{\bibinfo{volume}{103}}, \bibinfo{pages}{267001}
  (\bibinfo{year}{2009}).

\bibitem[{\citenamefont{Kontani et~al.}(2012)\citenamefont{Kontani, Inoue,
  Saito, Yamakawa, and Onari}}]{Kontani2012}
\bibinfo{author}{\bibfnamefont{H.}~\bibnamefont{Kontani}},
  \bibinfo{author}{\bibfnamefont{Y.}~\bibnamefont{Inoue}},
  \bibinfo{author}{\bibfnamefont{T.}~\bibnamefont{Saito}},
  \bibinfo{author}{\bibfnamefont{Y.}~\bibnamefont{Yamakawa}}, \bibnamefont{and}
  \bibinfo{author}{\bibfnamefont{S.}~\bibnamefont{Onari}},
  \bibinfo{journal}{Solid State Communications} \textbf{\bibinfo{volume}{152}},
  \bibinfo{pages}{718 } (\bibinfo{year}{2012}), ISSN \bibinfo{issn}{0038-1098},
  \bibinfo{note}{special Issue on Iron-based Superconductors}.

\bibitem[{\citenamefont{Onari and Kontani}(2012)}]{Onari2012}
\bibinfo{author}{\bibfnamefont{S.}~\bibnamefont{Onari}} \bibnamefont{and}
  \bibinfo{author}{\bibfnamefont{H.}~\bibnamefont{Kontani}},
  \bibinfo{journal}{Phys. Rev. Lett.} \textbf{\bibinfo{volume}{109}},
  \bibinfo{pages}{137001} (\bibinfo{year}{2012}).

\bibitem[{\citenamefont{Xu et~al.}(2008)\citenamefont{Xu, M\"uller, and
  Sachdev}}]{Xu2008}
\bibinfo{author}{\bibfnamefont{C.}~\bibnamefont{Xu}},
  \bibinfo{author}{\bibfnamefont{M.}~\bibnamefont{M\"uller}}, \bibnamefont{and}
  \bibinfo{author}{\bibfnamefont{S.}~\bibnamefont{Sachdev}},
  \bibinfo{journal}{Phys. Rev. B} \textbf{\bibinfo{volume}{78}},
  \bibinfo{pages}{020501} (\bibinfo{year}{2008}).

\bibitem[{\citenamefont{Fernandes et~al.}(2010)\citenamefont{Fernandes,
  VanBebber, Bhattacharya, Chandra, Keppens, Mandrus, McGuire, Sales, Sefat,
  and Schmalian}}]{Fernandes2010b}
\bibinfo{author}{\bibfnamefont{R.~M.} \bibnamefont{Fernandes}},
  \bibinfo{author}{\bibfnamefont{L.~H.} \bibnamefont{VanBebber}},
  \bibinfo{author}{\bibfnamefont{S.}~\bibnamefont{Bhattacharya}},
  \bibinfo{author}{\bibfnamefont{P.}~\bibnamefont{Chandra}},
  \bibinfo{author}{\bibfnamefont{V.}~\bibnamefont{Keppens}},
  \bibinfo{author}{\bibfnamefont{D.}~\bibnamefont{Mandrus}},
  \bibinfo{author}{\bibfnamefont{M.~A.} \bibnamefont{McGuire}},
  \bibinfo{author}{\bibfnamefont{B.~C.} \bibnamefont{Sales}},
  \bibinfo{author}{\bibfnamefont{A.~S.} \bibnamefont{Sefat}}, \bibnamefont{and}
  \bibinfo{author}{\bibfnamefont{J.}~\bibnamefont{Schmalian}},
  \bibinfo{journal}{Phys. Rev. Lett.} \textbf{\bibinfo{volume}{105}},
  \bibinfo{pages}{157003} (\bibinfo{year}{2010}).

\bibitem[{\citenamefont{Ning et~al.}(2010)\citenamefont{Ning, Ahilan, Imai,
  Sefat, McGuire, Sales, Mandrus, Cheng, Shen, and Wen}}]{Ning2010}
\bibinfo{author}{\bibfnamefont{F.~L.} \bibnamefont{Ning}},
  \bibinfo{author}{\bibfnamefont{K.}~\bibnamefont{Ahilan}},
  \bibinfo{author}{\bibfnamefont{T.}~\bibnamefont{Imai}},
  \bibinfo{author}{\bibfnamefont{A.~S.} \bibnamefont{Sefat}},
  \bibinfo{author}{\bibfnamefont{M.~A.} \bibnamefont{McGuire}},
  \bibinfo{author}{\bibfnamefont{B.~C.} \bibnamefont{Sales}},
  \bibinfo{author}{\bibfnamefont{D.}~\bibnamefont{Mandrus}},
  \bibinfo{author}{\bibfnamefont{P.}~\bibnamefont{Cheng}},
  \bibinfo{author}{\bibfnamefont{B.}~\bibnamefont{Shen}}, \bibnamefont{and}
  \bibinfo{author}{\bibfnamefont{H.-H.} \bibnamefont{Wen}},
  \bibinfo{journal}{Phys. Rev. Lett.} \textbf{\bibinfo{volume}{104}},
  \bibinfo{pages}{037001} (\bibinfo{year}{2010}).

\bibitem[{\citenamefont{Nakai et~al.}(2010)\citenamefont{Nakai, Iye, Kitagawa,
  Ishida, Ikeda, Kasahara, Shishido, Shibauchi, Matsuda, and
  Terashima}}]{Nakai2010}
\bibinfo{author}{\bibfnamefont{Y.}~\bibnamefont{Nakai}},
  \bibinfo{author}{\bibfnamefont{T.}~\bibnamefont{Iye}},
  \bibinfo{author}{\bibfnamefont{S.}~\bibnamefont{Kitagawa}},
  \bibinfo{author}{\bibfnamefont{K.}~\bibnamefont{Ishida}},
  \bibinfo{author}{\bibfnamefont{H.}~\bibnamefont{Ikeda}},
  \bibinfo{author}{\bibfnamefont{S.}~\bibnamefont{Kasahara}},
  \bibinfo{author}{\bibfnamefont{H.}~\bibnamefont{Shishido}},
  \bibinfo{author}{\bibfnamefont{T.}~\bibnamefont{Shibauchi}},
  \bibinfo{author}{\bibfnamefont{Y.}~\bibnamefont{Matsuda}}, \bibnamefont{and}
  \bibinfo{author}{\bibfnamefont{T.}~\bibnamefont{Terashima}},
  \bibinfo{journal}{Phys. Rev. Lett.} \textbf{\bibinfo{volume}{105}},
  \bibinfo{pages}{107003} (\bibinfo{year}{2010}).

\bibitem[{\citenamefont{Fernandes et~al.}(2013)\citenamefont{Fernandes,
  B\"ohmer, Meingast, and Schmalian}}]{Fernandes2013b}
\bibinfo{author}{\bibfnamefont{R.~M.} \bibnamefont{Fernandes}},
  \bibinfo{author}{\bibfnamefont{A.~E.} \bibnamefont{B\"ohmer}},
  \bibinfo{author}{\bibfnamefont{C.}~\bibnamefont{Meingast}}, \bibnamefont{and}
  \bibinfo{author}{\bibfnamefont{J.}~\bibnamefont{Schmalian}},
  \bibinfo{journal}{Phys. Rev. Lett.} \textbf{\bibinfo{volume}{111}},
  \bibinfo{pages}{137001} (\bibinfo{year}{2013}).

\bibitem[{\citenamefont{Gallais et~al.}(2013)\citenamefont{Gallais, Fernandes,
  Paul, Chauvi\`ere, Yang, M\'easson, Cazayous, Sacuto, Colson, and
  Forget}}]{Gallais2013}
\bibinfo{author}{\bibfnamefont{Y.}~\bibnamefont{Gallais}},
  \bibinfo{author}{\bibfnamefont{R.~M.} \bibnamefont{Fernandes}},
  \bibinfo{author}{\bibfnamefont{I.}~\bibnamefont{Paul}},
  \bibinfo{author}{\bibfnamefont{L.}~\bibnamefont{Chauvi\`ere}},
  \bibinfo{author}{\bibfnamefont{Y.-X.} \bibnamefont{Yang}},
  \bibinfo{author}{\bibfnamefont{M.-A.} \bibnamefont{M\'easson}},
  \bibinfo{author}{\bibfnamefont{M.}~\bibnamefont{Cazayous}},
  \bibinfo{author}{\bibfnamefont{A.}~\bibnamefont{Sacuto}},
  \bibinfo{author}{\bibfnamefont{D.}~\bibnamefont{Colson}}, \bibnamefont{and}
  \bibinfo{author}{\bibfnamefont{A.}~\bibnamefont{Forget}},
  \bibinfo{journal}{Phys. Rev. Lett.} \textbf{\bibinfo{volume}{111}},
  \bibinfo{pages}{267001} (\bibinfo{year}{2013}).

\bibitem[{\citenamefont{{Zhang} et~al.}(2014)\citenamefont{{Zhang}, {Richard},
  {Ding}, {Sefat}, {Gillett}, {Sebastian}, {Khodas}, and
  {Blumberg}}}]{Zhang2014}
\bibinfo{author}{\bibfnamefont{W.-L.} \bibnamefont{{Zhang}}},
  \bibinfo{author}{\bibfnamefont{P.}~\bibnamefont{{Richard}}},
  \bibinfo{author}{\bibfnamefont{H.}~\bibnamefont{{Ding}}},
  \bibinfo{author}{\bibfnamefont{A.~S.} \bibnamefont{{Sefat}}},
  \bibinfo{author}{\bibfnamefont{J.}~\bibnamefont{{Gillett}}},
  \bibinfo{author}{\bibfnamefont{S.~E.} \bibnamefont{{Sebastian}}},
  \bibinfo{author}{\bibfnamefont{M.}~\bibnamefont{{Khodas}}}, \bibnamefont{and}
  \bibinfo{author}{\bibfnamefont{G.}~\bibnamefont{{Blumberg}}},
  \bibinfo{journal}{ArXiv:1410.6452 e-prints}  (\bibinfo{year}{2014}),
  \bibinfo{note}{http://adsabs.harvard.edu/abs/2014arXiv1410.6452Z},
  \eprint{1410.6452}.

\bibitem[{\citenamefont{{Thorsm{\o}lle}
  et~al.}(2014)\citenamefont{{Thorsm{\o}lle}, {Khodas}, {Yin}, {Zhang}, {Carr},
  {Dai}, and {Blumberg}}}]{Thorsmolle2014}
\bibinfo{author}{\bibfnamefont{V.~K.} \bibnamefont{{Thorsm{\o}lle}}},
  \bibinfo{author}{\bibfnamefont{M.}~\bibnamefont{{Khodas}}},
  \bibinfo{author}{\bibfnamefont{Z.~P.} \bibnamefont{{Yin}}},
  \bibinfo{author}{\bibfnamefont{C.}~\bibnamefont{{Zhang}}},
  \bibinfo{author}{\bibfnamefont{S.~V.} \bibnamefont{{Carr}}},
  \bibinfo{author}{\bibfnamefont{P.}~\bibnamefont{{Dai}}}, \bibnamefont{and}
  \bibinfo{author}{\bibfnamefont{G.}~\bibnamefont{{Blumberg}}},
  \bibinfo{journal}{ArXiv:1410.6456 e-prints}  (\bibinfo{year}{2014}),
  \bibinfo{note}{http://adsabs.harvard.edu/abs/2014arXiv1410.6456T},
  \eprint{1410.6456}.

\bibitem[{\citenamefont{Klein and Dierker}(1984)}]{Klein1984}
\bibinfo{author}{\bibfnamefont{M.~V.} \bibnamefont{Klein}} \bibnamefont{and}
  \bibinfo{author}{\bibfnamefont{S.~B.} \bibnamefont{Dierker}},
  \bibinfo{journal}{Phys. Rev. B} \textbf{\bibinfo{volume}{29}},
  \bibinfo{pages}{4976} (\bibinfo{year}{1984}).

\bibitem[{\citenamefont{Devereaux and Hackl}(2007)}]{Devereaux2007}
\bibinfo{author}{\bibfnamefont{T.~P.} \bibnamefont{Devereaux}}
  \bibnamefont{and} \bibinfo{author}{\bibfnamefont{R.}~\bibnamefont{Hackl}},
  \bibinfo{journal}{Rev. Mod. Phys.} \textbf{\bibinfo{volume}{79}},
  \bibinfo{pages}{175} (\bibinfo{year}{2007}).

\bibitem[{\citenamefont{Cvetkovic and Vafek}(2013)}]{Cvetkovic2013}
\bibinfo{author}{\bibfnamefont{V.}~\bibnamefont{Cvetkovic}} \bibnamefont{and}
  \bibinfo{author}{\bibfnamefont{O.}~\bibnamefont{Vafek}},
  \bibinfo{journal}{Phys. Rev. B} \textbf{\bibinfo{volume}{88}},
  \bibinfo{pages}{134510} (\bibinfo{year}{2013}).

\bibitem[{\citenamefont{Luttinger}(1956)}]{Luttinger1956}
\bibinfo{author}{\bibfnamefont{J.~M.} \bibnamefont{Luttinger}},
  \bibinfo{journal}{Phys. Rev.} \textbf{\bibinfo{volume}{102}},
  \bibinfo{pages}{1030} (\bibinfo{year}{1956}).

\bibitem[{\citenamefont{Kuroki et~al.}(2008)\citenamefont{Kuroki, Onari, Arita,
  Usui, Tanaka, Kontani, and Aoki}}]{Kuroki2008}
\bibinfo{author}{\bibfnamefont{K.}~\bibnamefont{Kuroki}},
  \bibinfo{author}{\bibfnamefont{S.}~\bibnamefont{Onari}},
  \bibinfo{author}{\bibfnamefont{R.}~\bibnamefont{Arita}},
  \bibinfo{author}{\bibfnamefont{H.}~\bibnamefont{Usui}},
  \bibinfo{author}{\bibfnamefont{Y.}~\bibnamefont{Tanaka}},
  \bibinfo{author}{\bibfnamefont{H.}~\bibnamefont{Kontani}}, \bibnamefont{and}
  \bibinfo{author}{\bibfnamefont{H.}~\bibnamefont{Aoki}},
  \bibinfo{journal}{Phys. Rev. Lett.} \textbf{\bibinfo{volume}{101}},
  \bibinfo{pages}{087004} (\bibinfo{year}{2008}).

\bibitem[{\citenamefont{Cvetkovic and Tesanovic}(2009)}]{Cvetkovic2009}
\bibinfo{author}{\bibfnamefont{V.}~\bibnamefont{Cvetkovic}} \bibnamefont{and}
  \bibinfo{author}{\bibfnamefont{Z.}~\bibnamefont{Tesanovic}},
  \bibinfo{journal}{EPL (Europhysics Letters)} \textbf{\bibinfo{volume}{85}},
  \bibinfo{pages}{37002} (\bibinfo{year}{2009}).

\bibitem[{\citenamefont{Li et~al.}(2009)\citenamefont{Li, Chen, Chang, Lynn,
  Li, Luo, Cao, Xu, and Dai}}]{Li2009}
\bibinfo{author}{\bibfnamefont{S.}~\bibnamefont{Li}},
  \bibinfo{author}{\bibfnamefont{Y.}~\bibnamefont{Chen}},
  \bibinfo{author}{\bibfnamefont{S.}~\bibnamefont{Chang}},
  \bibinfo{author}{\bibfnamefont{J.~W.} \bibnamefont{Lynn}},
  \bibinfo{author}{\bibfnamefont{L.}~\bibnamefont{Li}},
  \bibinfo{author}{\bibfnamefont{Y.}~\bibnamefont{Luo}},
  \bibinfo{author}{\bibfnamefont{G.}~\bibnamefont{Cao}},
  \bibinfo{author}{\bibfnamefont{Z.}~\bibnamefont{Xu}}, \bibnamefont{and}
  \bibinfo{author}{\bibfnamefont{P.}~\bibnamefont{Dai}},
  \bibinfo{journal}{Phys. Rev. B} \textbf{\bibinfo{volume}{79}},
  \bibinfo{pages}{174527} (\bibinfo{year}{2009}).

\bibitem[{\citenamefont{Tucker et~al.}(2012)\citenamefont{Tucker, Pratt, Kim,
  Ran, Thaler, Granroth, Marty, Tian, Zarestky, Lumsden, Bud'ko, Canfield,
  Kreyssig, Goldman, and McQueeney}}]{Tucker2012}
\bibinfo{author}{\bibfnamefont{G.~S.} \bibnamefont{Tucker}},
  \bibinfo{author}{\bibfnamefont{D.~K.} \bibnamefont{Pratt}},
  \bibinfo{author}{\bibfnamefont{M.~G.} \bibnamefont{Kim}},
  \bibinfo{author}{\bibfnamefont{S.}~\bibnamefont{Ran}},
  \bibinfo{author}{\bibfnamefont{A.}~\bibnamefont{Thaler}},
  \bibinfo{author}{\bibfnamefont{G.~E.} \bibnamefont{Granroth}},
  \bibinfo{author}{\bibfnamefont{K.}~\bibnamefont{Marty}},
  \bibinfo{author}{\bibfnamefont{W.}~\bibnamefont{Tian}},
  \bibinfo{author}{\bibfnamefont{J.~L.} \bibnamefont{Zarestky}},
  \bibinfo{author}{\bibfnamefont{M.~D.} \bibnamefont{Lumsden}},
  \bibinfo{author}{\bibfnamefont{S.~L.} \bibnamefont{Bud'ko}},
  \bibinfo{author}{\bibfnamefont{P.~C.} \bibnamefont{Canfield}},
  \bibinfo{author}{\bibfnamefont{A.}~\bibnamefont{Kreyssig}},
  \bibinfo{author}{\bibfnamefont{A.~I.} \bibnamefont{Goldman}},
  \bibnamefont{and} \bibinfo{author}{\bibfnamefont{R.~J.}
  \bibnamefont{McQueeney}}, \bibinfo{journal}{Phys. Rev. B}
  \textbf{\bibinfo{volume}{86}}, \bibinfo{pages}{020503}
  (\bibinfo{year}{2012}).

\bibitem[{\citenamefont{Paul}(2014)}]{Paul2014}
\bibinfo{author}{\bibfnamefont{I.}~\bibnamefont{Paul}}, \bibinfo{journal}{Phys.
  Rev. B} \textbf{\bibinfo{volume}{90}}, \bibinfo{pages}{115102}
  (\bibinfo{year}{2014}).

\bibitem[{\citenamefont{{Gallais} et~al.}(2015)\citenamefont{{Gallais}, {Paul},
  {Chauviere}, and {Schmalian}}}]{Gallais2015}
\bibinfo{author}{\bibfnamefont{Y.}~\bibnamefont{{Gallais}}},
  \bibinfo{author}{\bibfnamefont{I.}~\bibnamefont{{Paul}}},
  \bibinfo{author}{\bibfnamefont{L.}~\bibnamefont{{Chauviere}}},
  \bibnamefont{and}
  \bibinfo{author}{\bibfnamefont{J.}~\bibnamefont{{Schmalian}}},
  \bibinfo{journal}{ArXiv e-prints}  (\bibinfo{year}{2015}),
  \eprint{1504.04570}.

\bibitem[{\citenamefont{{Karahasanovic}
  et~al.}(2015)\citenamefont{{Karahasanovic}, {Kretzschmar}, {Boehm}, {Hackl},
  {Paul}, {Gallais}, and {Schmalian}}}]{Karahasanovic2015}
\bibinfo{author}{\bibfnamefont{U.}~\bibnamefont{{Karahasanovic}}},
  \bibinfo{author}{\bibfnamefont{F.}~\bibnamefont{{Kretzschmar}}},
  \bibinfo{author}{\bibfnamefont{T.}~\bibnamefont{{Boehm}}},
  \bibinfo{author}{\bibfnamefont{R.}~\bibnamefont{{Hackl}}},
  \bibinfo{author}{\bibfnamefont{I.}~\bibnamefont{{Paul}}},
  \bibinfo{author}{\bibfnamefont{Y.}~\bibnamefont{{Gallais}}},
  \bibnamefont{and}
  \bibinfo{author}{\bibfnamefont{J.}~\bibnamefont{{Schmalian}}},
  \bibinfo{journal}{ArXiv e-prints}  (\bibinfo{year}{2015}),
  \eprint{1504.06841}.

\bibitem[{\citenamefont{Kontani and Yamakawa}(2014)}]{Kontani2014}
\bibinfo{author}{\bibfnamefont{H.}~\bibnamefont{Kontani}} \bibnamefont{and}
  \bibinfo{author}{\bibfnamefont{Y.}~\bibnamefont{Yamakawa}},
  \bibinfo{journal}{Phys. Rev. Lett.} \textbf{\bibinfo{volume}{113}},
  \bibinfo{pages}{047001} (\bibinfo{year}{2014}).

\bibitem[{\citenamefont{Boehmer}(2015)}]{Boehmer2015}
\bibinfo{author}{\bibfnamefont{A.~E.} \bibnamefont{Boehmer}},
\bibinfo{author}{\bibfnamefont{C.} \bibnamefont{Meingast}},
  \bibinfo{journal}{ArXiv:1505.05120}.


\bibitem[{\citenamefont{Platzman}(1965)}]{Platzman1965}
\bibinfo{author}{\bibfnamefont{P.~M.} \bibnamefont{Platzman}},
  \bibinfo{journal}{Phys. Rev.} \textbf{\bibinfo{volume}{139}},
  \bibinfo{pages}{A379} (\bibinfo{year}{1965}).

\bibitem[{\citenamefont{Yamase and Zeyher}(2013)}]{Yamase2013}
\bibinfo{author}{\bibfnamefont{H.}~\bibnamefont{Yamase}} \bibnamefont{and}
  \bibinfo{author}{\bibfnamefont{R.}~\bibnamefont{Zeyher}},
  \bibinfo{journal}{Phys. Rev. B} \textbf{\bibinfo{volume}{88}},
  \bibinfo{pages}{125120} (\bibinfo{year}{2013}).

\bibitem[{\citenamefont{Caprara et~al.}(2005)\citenamefont{Caprara, Di~Castro,
  Grilli, and Suppa}}]{Caprara2005}
\bibinfo{author}{\bibfnamefont{S.}~\bibnamefont{Caprara}},
  \bibinfo{author}{\bibfnamefont{C.}~\bibnamefont{Di~Castro}},
  \bibinfo{author}{\bibfnamefont{M.}~\bibnamefont{Grilli}}, \bibnamefont{and}
  \bibinfo{author}{\bibfnamefont{D.}~\bibnamefont{Suppa}},
  \bibinfo{journal}{Phys. Rev. Lett.} \textbf{\bibinfo{volume}{95}},
  \bibinfo{pages}{117004} (\bibinfo{year}{2005}).

\end{thebibliography}
\end{document}